\RequirePackage{ifpdf}
\ifpdf 
\documentclass[pdftex]{sigma}
\else
\documentclass{sigma}
\fi

\begin{document}
\allowdisplaybreaks

\renewcommand{\PaperNumber}{004}

\FirstPageHeading

\ShortArticleName{On Linearizing Systems of Diffusion Equations}

\ArticleName{On Linearizing Systems of Diffusion Equations}

\Author{Christodoulos SOPHOCLEOUS~$^\dag$ and Ron J. WILTSHIRE~$^\ddag$}
\AuthorNameForHeading{C.~Sophocleous  and R.J.~Wiltshire}

\Address{$^\dag$~Department of Mathematics and Statistics,
University of Cyprus, CY 1678 Nicosia, Cyprus}
\EmailD{\href{mailto:christod@ucy.ac.cy}{christod@ucy.ac.cy}} 
\URLaddressD{\url{http://www.ucy.ac.cy/~christod/}} 

\Address{$^\ddag$~The Division of Mathematics and Statistics,
The University of Glamorgan,\\
$\phantom{^\ddag}$~Pontypridd CF37 1DL, Great Britain}
\EmailD{\href{mailto:rjwiltsh@glam.ac.uk}{rjwiltsh@glam.ac.uk}}

\ArticleDates{Received November 23, 2005, in final form January 10,
2006; Published online January 16, 2006}

\Abstract{We consider  systems of diffusion equations
that have considerable
interest in Soil Science and Mathematical Biology and focus upon the problem of finding those
forms of this class that can be linearized. In particular we use the equivalence transformations of the
second generation potential system to derive forms of this system that can be linearized.
In turn, these transformations lead to nonlocal mappings that linearize the original system.}


\Classification{35A30; 58J70; 58J72; 92B05}

\section{Introduction}

Whilst systems of pure diffusion equations, in both their linear
and nonlinear forms are well known and have many physical and
biological applications, the research described here focusses on
less familiar cases where diffusion coefficients or other `shape'
functions are defined either in general or poor analytic terms. Of
particular interest here is the case of the extension of Richard's
equation, which describes the movement of water in a homogeneous
unsaturated soil, to cases describing the combined transport of
water vapour and heat under a combination of gradients of soil
temperature and volumetric water content. Such coupled transport
is of considerable significance in semi-arid environments where
moisture transport often occurs essentially in the water vapour
phase \cite{sw:philip}. Under these conditions the transport
equations, valid in a vertical column of soil, may be written in
the form \cite{sw:jury}
\begin{gather}
\frac{\partial u}{\partial t}=\frac{\partial~}{\partial x}\left
[f(u,v)\frac{\partial u}{\partial x}+g(u,v)\frac{\partial v}{\partial
x}\right ], \nonumber \\
\frac{\partial v}{\partial t}=\frac{\partial~}{\partial x}\left
[h(u,v)\frac{\partial u}{\partial x}+k(u,v)\frac{\partial
v}{\partial x}\right ],\label{sw:a1}
\end{gather}
where $u(x, t)$ and $v(x, t)$ are respectively the soil
temperature and volumetric water content at depth $x$ and time
$t$. It is important to realize that extensions which include the
coupled diffusion of solute follow in an obvious way.

If we introduce the potential variable $w$, we can generate the auxiliary system of (\ref{sw:a1}),
\begin{gather}
w_x=u, \nonumber\\
w_t=f(u,v)u_x+g(u,v)v_x,  \nonumber\\
v_t=\left [h(u,v)u_x+k(u,v)v_x\right ]_x.\label{sw:a2} 
\end{gather}
Introduce potential variables $w$ and $z$ we generate the second generation auxiliary system of (\ref{sw:a1}),
\begin{gather}
w_x=u, \nonumber \\
w_t=f(u,v)u_x+g(u,v)v_x, \nonumber \\
z_x=v,  \nonumber \\
z_t=h(u,v)u_x+k(u,v)v_x. \label{sw:a3}
\end{gather}
The study of Lie symmetries of system (\ref{sw:a1}) has been considered in~\cite{sw:ron1,sw:ron2}.
It is known that Lie symmetries of the auxiliary systems, in some cases, lead to non-local symmetries known
as potential symmetries, for the original system. Lie symmetries for the systems  (\ref{sw:a2})
and~(\ref{sw:a3}) that induce potential symmetries for the original system (\ref{sw:a1}) have been considered
in~\cite{sw:sw}.

Here we calculate the equivalence transformations for the systems (\ref{sw:a1})--(\ref{sw:a3}). We use
the equi\-va\-lence transformations of (\ref{sw:a3}) to derive those forms of (\ref{sw:a3}) that can be linearized.
Consequently these point transformations lead to contact transformations that linearize the corresponding
forms of (\ref{sw:a1}). Furthermore for one case of (\ref{sw:a2}), which admits infinite-dimensional Lie
symmetries, we construct a linearizing mapping.

\section{Equivalence transformations}

An equivalence transformation is a nondegenerate change of the independent and dependent variables
taking a PDE into another PDE of the same form. For example, an equivalence transformation of the system
(\ref{sw:a1}) will transform it into a system of the same form, but in general, with different
functions $f(u,v)$, $g(u,v)$, $h(u,v)$ and $k(u,v)$. The set of all equivalence transformations forms
an equivalence group. We use the infinitesimal method \cite{sw:ovsiannikov} to determine the
equivalence transformations of the systems (\ref{sw:a1})--(\ref{sw:a3}).

\subsection{Equivalence transformations for system (\ref{sw:a1})}

We seek for equivalence group generators of the form
\[
X^E=\xi_1\frac{\partial~}{\partial x}+\xi_2\frac{\partial~}{\partial t}+
\eta_1\frac{\partial~}{\partial u}+\eta_2\frac{\partial~}{\partial v}+
\phi_1\frac{\partial~}{\partial f}+\phi_2\frac{\partial~}{\partial g}+
\phi_3\frac{\partial~}{\partial h}+\phi_4\frac{\partial~}{\partial k}.
\]
We will present the results, without giving any detailed calculations. The method for determining equivalence
transformations is presented in \cite{sw:ovsiannikov}. (See also in \cite{sw:ibragimov}). We find that
the system (\ref{sw:a1}) has a continuous group of equivalence transformations generated by the following
10 infinitesimal operators:
\begin{gather*}
X_1^E=\frac{\partial~}{\partial x},\qquad
X_2^E=\frac{\partial~}{\partial t},\qquad
X_3^E=\frac{\partial~}{\partial u},\qquad
X_4^E=\frac{\partial~}{\partial v}, \\
X_5^E=x\frac{\partial~}{\partial x}+2f\frac{\partial~}{\partial f}+2g\frac{\partial~}{\partial g}+
2h\frac{\partial~}{\partial h}+2k\frac{\partial~}{\partial k}, \\
X_6^E=t\frac{\partial~}{\partial t}-f\frac{\partial~}{\partial f}-g\frac{\partial~}{\partial g}-
h\frac{\partial~}{\partial h}-k\frac{\partial~}{\partial k}, \qquad
X_7^E=u\frac{\partial~}{\partial u}+g\frac{\partial~}{\partial g}, \\
X_8^E=v\frac{\partial~}{\partial v}-g\frac{\partial~}{\partial g}+h\frac{\partial~}{\partial h}, \qquad
X_9^E=v\frac{\partial~}{\partial u}+h\frac{\partial~}{\partial f}+(k-f)\frac{\partial~}{\partial g}-
h\frac{\partial~}{\partial k}, \\
X_{10}^E=u\frac{\partial~}{\partial v}-g\frac{\partial~}{\partial f}+(f-k)\frac{\partial~}{\partial h}+
g\frac{\partial~}{\partial k}.
\end{gather*}

Using Lie's theorem we show that the above equivalence transformations in finite form read
\[
x^{\prime}=c_1x+c_2, \qquad t^{\prime}=c_3t+c_4,\qquad u^{\prime}=c_5u+c_6v+c_7,\qquad v^{\prime}=c_8u+c_9v+c_{10},
\]
where
\begin{gather*}
f^{\prime}=\frac{c_1^2\left [c_5(c_9f-c_8g)+c_6(c_9h-c_8k)\right ]}{c_3(c_5c_9-c_6c_8)},\qquad
g^{\prime}=\frac{c_1^2\left [-c_5(c_6f-c_5g)-c_6(c_6h-c_5k)\right ]}{c_3(c_5c_9-c_6c_8)}, \\
h^{\prime}=\frac{c_1^2\left [c_8(c_9f-c_8g)+c_9(c_9h-c_8k)\right ]}{c_3(c_5c_9-c_6c_8)},\qquad
k^{\prime}=\frac{c_1^2\left [-c_8(c_6f-c_5g)-c_9(c_6h-c_5k)\right ]}{c_3(c_5c_9-c_6c_8)}.
\end{gather*}

Clearly, if the functions $f$, $g$, $h$, $k$ are linearly dependent, then one of the functions of the transformed
equation can be taken equal to zero, provided that the constants involved satisfy certain
relations. That is, system (\ref{sw:a1}) can  be mapped into a system of the same form
but with one of the functions to be zero. For example, if $f=g-h+k$ then the mapping
$x\mapsto x$, $t\mapsto t$, $u\mapsto u+v$, $v\mapsto u+cv$, $c\ne 1$ transforms the system~(\ref{sw:a1})
into a system of the same form, but with $g=0$.

\subsection{Equivalence transformations for system (\ref{sw:a2})}

We find that
the system (\ref{sw:a2}) has a continuous group of equivalence transformations generated by the following
9 infinitesimal operators:
\begin{gather*}
Y_1^E=\frac{\partial~}{\partial x},\qquad
Y_2^E=\frac{\partial~}{\partial t},\qquad
Y_3^E=\frac{\partial~}{\partial u}+x\frac{\partial~}{\partial w},~~
Y_4^E=\frac{\partial~}{\partial v},\qquad
Y_5^E=\frac{\partial~}{\partial w},\\
Y_6^E=u\frac{\partial~}{\partial u}+v\frac{\partial~}{\partial v}+w\frac{\partial~}{\partial w}, \qquad
Y_7^E=x\frac{\partial~}{\partial x}+w\frac{\partial~}{\partial w}+2f\frac{\partial~}{\partial f}+
2g\frac{\partial~}{\partial g}+2h\frac{\partial~}{\partial h}+2k\frac{\partial~}{\partial k}, \\
Y_8^E=t\frac{\partial~}{\partial t}-f\frac{\partial~}{\partial f}-g\frac{\partial~}{\partial g}-
h\frac{\partial~}{\partial h}-k\frac{\partial~}{\partial k}, \qquad
Y_9^E=v\frac{\partial~}{\partial v}-g\frac{\partial~}{\partial g}+h\frac{\partial~}{\partial h}.
\end{gather*}

The equivalence transformations, in finite form, read
\begin{gather*}
x^{\prime}=c_1x+c_2,\qquad t^{\prime}=c_3t+c_4,\qquad u^{\prime}=c_5u+c_6,\qquad v^{\prime}=c_7v+c_{8},\\
w^{\prime}=c_1c_5w+c_1c_6x+c_9,
\end{gather*}
where
\[
f^{\prime}=\frac{c_1^2f}{c_3},\qquad g^{\prime}=\frac{c_5c_1^2g}{c_7c_3},\qquad
h^{\prime}=\frac{c_7c_1^2h}{c_5c_3},\qquad k^{\prime}=\frac{c_1^2k}{c_3}.
\]

\subsection{Equivalence transformations for system (\ref{sw:a3})}

We find that
the system (\ref{sw:a3}) has a continuous group of equivalence transformations generated by the following
14 infinitesimal operators:
\begin{gather*}
Z_{1}^{E}=\frac{\partial~}{\partial x},\qquad 
Z_{2}^{E}=\frac{\partial~}{\partial t},\qquad
Z_{3}^{E}=\frac{\partial~}{\partial w},\qquad
Z_{4}^{E}=\frac{\partial~}{\partial z}, \\
Z_{5}^{E}=z\frac{\partial~}{\partial x}-uv\frac
{\partial~}{\partial u}-v^{2}\frac{\partial~}{\partial v}+\left(  2vf-uh\right)
\frac{\partial~}{\partial f}\\
\phantom{Z_{5}^{E}=}{}
+\left(  -uk+3vg+uf\right)  \frac{\partial~}{\partial
g}+vh\frac{\partial~}{\partial h}+\left(  2vk+uh\right)  \frac{\partial~}{\partial k}, \\
Z_{6}^{E}=w\frac{\partial~}{\partial x}-u^{2}
\frac{\partial~}{\partial u}-uv\frac{\partial~}{\partial v}+\left(
vg+2uf\right)  \frac{\partial~}{\partial f}\\
\phantom{Z_{6}^{E}=}{}
+ug\frac{\partial~}{\partial g}+\left(
vk+3uh-vf\right)  \frac{\partial~}{\partial h}+\left(  2uk-vg\right)
\frac{\partial~}{\partial k}, \\
Z_{7}^{E}=x\frac{\partial~}{\partial x}-u\frac{\partial~}
{\partial u}-v\frac{\partial~}{\partial v}+2f\frac{\partial~}{\partial
f}+2g\frac{\partial~}{\partial g}+2h\frac{\partial~}{\partial h}+2k\frac
{\partial~}{\partial k}, \\
Z_{8}^{E}=t\frac{\partial~}{\partial t}-f\frac{\partial~}
{\partial f}-g\frac{\partial~}{\partial g}-h\frac{\partial~}{\partial h}
-k\frac{\partial~}{\partial k}, \\
Z_{9}^{E}=v\frac{\partial~}{\partial u}+z\frac{\partial~}
{\partial w}+h\frac{\partial~}{\partial f}+\left(  k-f\right)  \frac{\partial~}
{\partial g}-h\frac{\partial~}{\partial k}, \\
Z_{10}^{E}=u\frac{\partial~}{\partial v}+w\frac
{\partial~}{\partial z}-g\frac{\partial~}{\partial f}+\left(  f-k\right)
\frac{\partial~}{\partial h}+g\frac{\partial~}{\partial k}, \qquad
Z_{11}^{E}=v\frac{\partial~}{\partial v}+z\frac
{\partial~}{\partial z}-g\frac{\partial~}{\partial g}+h\frac{\partial~}{\partial h}, \\
Z_{12}^{E}=u\frac{\partial~}{\partial u}+w\frac
{\partial~}{\partial w}+g\frac{\partial~}{\partial g}-h\frac{\partial~}{\partial h}, \qquad
Z_{13}^{E}=\frac{\partial~}{\partial v}+x\frac{\partial}{\partial z}, \qquad
Z_{14}^{E}=\frac{\partial~}{\partial u}+x\frac{\partial~}{\partial w}.
\end{gather*}

Naturally using Lie's theorem we can find the equivalence transformations in finite form. We find that
\begin{gather*}
x^{\prime}=ax +p_1w+p_2z+c_1,\qquad t^{\prime}=\frac{1}{\gamma}t+\delta,
\qquad w^{\prime}=q_1x+a_1w+a_2z+c_2,\\ z^{\prime}=q_2x+a_3w+a_4z+c_3.
\end{gather*}
However the forms of $u^{\prime}$ and $v^{\prime}$ cannot be found easily from Lie's theorem. In the next section
we show that they take the form
\[
u^{\prime}=\frac{q_1+a_1u+a_2v}{a+p_1u+p_2v},\qquad
v^{\prime}=\frac{q_2+a_3u+a_4v}{a+p_1u+p_2v} .
\]
Furthermore we state that the forms of $f^{\prime}$, $g^{\prime}$, $h^{\prime}$ and $k^{\prime}$, which are all linear
in $f$, $g$, $h$ and $k$, are very lengthy.

In the following section we use the finite form of the equivalence transformations to
derive special cases of the system~(\ref{sw:a3}) that can be linearized. In turn these transformations
lead to nonlocal mappings that linearize the corresponding forms~(\ref{sw:a1}).

\section{On linearization}

Here we consider the problem of finding forms of the nonlinear system (\ref{sw:a3}) that can be linearized.
We adopt the idea of the transformation
\begin{gather}\label{sw:cc1}
x^{\prime}=v,\qquad t^{\prime}=t,\qquad u^{\prime}=\frac 1u,\qquad v^{\prime}=x
\end{gather}
that maps the auxiliary system of the nonlinear diffusion equation $u_t=[u^{-2}u_x]_x$ \cite{sw:bluman2},
\[
v_x=u,\qquad v_t=u^{-2}u_x
\]
into the auxiliary system of the linear diffusion equation $u^{\prime}_{t^{\prime}}=u^{\prime}_{x^{\prime}x^{\prime}}$,
\[
v^{\prime}_{x^{\prime}}=u^{\prime},\qquad v^{\prime}_{t^{\prime}}=u^{\prime}_{x^{\prime}} .
\]
The above transformation is a member of the equivalence transformations of
\[
v_x=u,\qquad v_t=f(u)u_x.
\]
Motivated by the above results we find the forms of $f(u,v)$, $g(u,v)$, $h(u,v)$ and $k(u,v)$ such that
system (\ref{sw:a3}) can be mapped into a linear system by the equivalence transformations admitted 
by~(\ref{sw:a3}). These local mappings will lead to nonlocal mappings that linearize the corresponding forms
of the system~(\ref{sw:a1}).

We consider system~(\ref{sw:a1}) with the four functions equal to constants. That is, it
takes the linear form
\begin{gather}
u^{\prime}_{t^{\prime}}=\mu_1u^{\prime}_{x^{\prime}x^{\prime}}+\mu_2v^{\prime}_{x^{\prime}x^{\prime}},  \qquad
v^{\prime}_{t^{\prime}}=\mu_3u^{\prime}_{x^{\prime}x^{\prime}}+\mu_4v^{\prime}_{x^{\prime}x^{\prime}}.\label{sw:c1}
\end{gather}
We introduce the potential variables $w^{\prime}$ and $z^{\prime}$ to obtain the auxiliary system
of (\ref{sw:c1})
\begin{gather}
w^{\prime}_{x^{\prime}}=u^{\prime}, \qquad
w^{\prime}_{t^{\prime}}=\mu_1u^{\prime}_{x^{\prime}}+\mu_2v^{\prime}_{x^{\prime}}, \qquad
z^{\prime}_{x^{\prime}}=v^{\prime},  \qquad
z^{\prime}_{t^{\prime}}=\mu_3u^{\prime}_{x^{\prime}}+\mu_4v^{\prime}_{x^{\prime}}. \label{sw:c2}
\end{gather}

From the equivalence transformations of the system (\ref{sw:a3}) we deduce the finite transformations
\begin{gather}\label{sw:c3}
x^{\prime}=ax +p_1w+p_2z,\qquad w^{\prime}=q_1x+a_1w+a_2z,\qquad z^{\prime}=q_2x+a_3w+a_4z
\end{gather}
where we have taken, without loss of generality, the translation constants equal to zero. Clearly,
the inverse transformations is of the form
\begin{gather}\label{sw:c4}
x=bx^{\prime}+r_1w^{\prime}+r_2z^{\prime},\qquad
w=s_1x^{\prime}+b_1w^{\prime}+b_2z^{\prime},\qquad
z=s_2x^{\prime}+b_3w^{\prime}+b_4z^{\prime}.
\end{gather}
Also from the equivalence transformations we deduce that
\[
t=\gamma t^{\prime},
\]
with the translation constant taken to be zero.

Introducing vector notation, we can write  systems (\ref{sw:a3}) and (\ref{sw:c2}) in the form
\begin{gather}\label{sw:c5}
{\boldsymbol w}_x={\boldsymbol u},\qquad {\boldsymbol w}_t=F({\boldsymbol u} ){\boldsymbol u}_x
\end{gather}
and
\begin{gather}\label{sw:c6}
{\boldsymbol w}^{\prime}_{x^{\prime}}={\boldsymbol u^{\prime}},\qquad
{\boldsymbol w}^{\prime}_{t^{\prime}}=\Lambda{\boldsymbol u^{\prime}}_{x^{\prime}}
\end{gather}
respectively, where
\[
{\boldsymbol w}=\left [\begin{array}{c}
w\\z\end{array} \right ],\qquad
{\boldsymbol u}=\left [\begin{array}{c}
u\\v \end{array}\right ],\qquad F({\boldsymbol u})=\left [\begin{array}{cc}
f(u,v)&g(u,v)\\h(u,v)&k(u,v)\end{array}\right ],\qquad \Lambda =\left [\begin{array}{cc}
\mu_1&\mu_2\\\mu_3&\mu_4\end{array}\right ].
\]
Furthermore transformation (\ref{sw:c3}) can be written in the form
\begin{gather}\label{sw:c7}
\left[
\begin{array}
[c]{c}
x^{\prime}\\
{\boldsymbol w}^{\prime}
\end{array}
\right]  =\left[
\begin{array}
[c]{cc}
a & {\boldsymbol p}^{T}\\
{\boldsymbol q} &  A
\end{array}
\right]  \left[
\begin{array}
[c]{c}
x\\
{\boldsymbol w}
\end{array}
\right],
\end{gather}
where
\[
{\boldsymbol p}=\left [\begin{array}{c}
p_1\\p_2\end{array} \right ],\qquad {\boldsymbol q}=\left [\begin{array}{c}
q_1\\q_2\end{array} \right ],\qquad
A=\left [\begin{array}{cc}
a_1&a_2\\a_3&a_4 \end{array} \right ]
\]
and the inverse transformation (\ref{sw:c4}) takes the form
\begin{gather}\label{sw:c8}
\left[
\begin{array}
[c]{c}
x\\
{\boldsymbol w}
\end{array}
\right]  =\left[
\begin{array}
[c]{cc}
b & {\boldsymbol r}^{T}\\
{\boldsymbol s} &  B
\end{array}
\right]  \left[
\begin{array}
[c]{c}
x^{\prime}\\
{\boldsymbol w}^{\prime}
\end{array}
\right],
\end{gather}
where
\[
{\boldsymbol r}=\left [\begin{array}{c}
r_1\\r_2\end{array} \right ],\qquad {\boldsymbol s}=\left [\begin{array}{c}
s_1\\s_2\end{array} \right ],\qquad 
B=\left [\begin{array}{cc}
b_1&b_2\\b_3&b_4 \end{array} \right ].
\]
Therefore from the transformations ({\ref{sw:c7}) and (\ref{sw:c8}) we deduce that
\begin{gather}\label{sw:c9}
\left[
\begin{array}[c]{cc}
a & {\boldsymbol p}^{T}\\
{\boldsymbol q} & A
\end{array}
\right]  \left[
\begin{array}
[c]{cc}%
b & {\boldsymbol r}^{T}\\
{\boldsymbol s} & B
\end{array}
\right]  =\left[
\begin{array}
[c]{cc}%
ab+{\boldsymbol p\cdot s} & a{\boldsymbol r}^{T}+{\boldsymbol p}^{T}B\\
{\boldsymbol q}b+A{\boldsymbol s} & {\boldsymbol qr}^{T}+AB
\end{array}
\right]  =\left[
\begin{array}
[c]{cc}%
1 & 0\\
0 & I_{2}%
\end{array}
\right],
\end{gather}
where $I_{2}$ is the $2\times 2$ identity matrix. In addition:%
\begin{gather}\label{sw:c10}
\left[
\begin{array}
[c]{cc}%
b & {\boldsymbol r}^{T}\\
{\boldsymbol s} & B
\end{array}
\right]  \left[
\begin{array}
[c]{cc}%
a & {\boldsymbol p}^{T}\\
{\boldsymbol q} & A
\end{array}
\right]  =\left[
\begin{array}
[c]{cc}%
ab+{\boldsymbol r\cdot q} & b{\boldsymbol p}^{T}+{\boldsymbol r}^{T}A\\
{\boldsymbol s}a+B{\boldsymbol q} & {\boldsymbol sp}^{T}+BA
\end{array}
\right]  =\left[
\begin{array}
[c]{cc}%
1 & 0\\
0 & I_{2}%
\end{array}
\right] .
\end{gather}

Using normal transformation rules:
\[
{\boldsymbol w}_{x}={\boldsymbol w}_{x^{\prime}}\frac{\partial x^{\prime}}{\partial
x}+{\boldsymbol w}_{t^{\prime}}\frac{\partial t^{\prime}}{\partial x}
\]
and from the transformations (\ref{sw:c7}) and (\ref{sw:c8}) becomes
\[
{\boldsymbol w}_{x}=\left(  {\boldsymbol s}+B{\boldsymbol w}^{\prime}_{x^{\prime}}\right)
\left(  a+{\boldsymbol p\cdot w}_{x}\right) .
\]
Thus from the two potential systems (\ref{sw:c5}) and (\ref{sw:c6})
\[
{\boldsymbol u}=\left(  {\boldsymbol s}+B{\boldsymbol u^{\prime}}\right)
\left(  a+{\boldsymbol p\cdot u}\right) .
\]
Finally, from the relations (\ref{sw:c9}) and (\ref{sw:c10}) we obtain
\[
{\boldsymbol u^{\prime}}=\frac{A{\boldsymbol u+ q}}{a+{\boldsymbol p\cdot u}} .
\]
Hence,
\[
u^{\prime}=\frac{q_1+a_1u+a_2v}{a+p_1u+p_2v},\qquad
v^{\prime}=\frac{q_2+a_3u+a_4v}{a+p_1u+p_2v} .
\]
These forms of $u^{\prime}$ and $v^{\prime}$ can also be obtained from the
equivalence transformations, however, in a more complicated manner.

Consider now
\[
{\boldsymbol w}^{\prime}_{t^{\prime}}={\boldsymbol w}^{\prime}_x\frac{\partial x}
{\partial t^{\prime}}+{\boldsymbol w}^{\prime}_{t}\frac{\partial t}
{\partial t^{\prime}}
\]
and from the transformations (\ref{sw:c7}) and (\ref{sw:c8}) and the two potential systems
(\ref{sw:c5}) and (\ref{sw:c6}) we find
\begin{gather}\label{sw:c11}
\Lambda {\boldsymbol u}^{\prime}_{x^{\prime}}=({\boldsymbol q}+A{\boldsymbol u})({\boldsymbol r}^T\Lambda {\boldsymbol u}^{\prime}_{x^{\prime}})
+\gamma AF{\boldsymbol u}_x .
\end{gather}
Multiply by ${\boldsymbol r}^T$ and using the relations (\ref{sw:c9}) and (\ref{sw:c10}) it follows that
\[
{\boldsymbol r}^T\Lambda {\boldsymbol u}^{\prime}_{x^{\prime}}=-\frac{\gamma {\boldsymbol p}^TF{\boldsymbol u}_x}{a+{\boldsymbol p\cdot u}}
\]
and also consider
\[
{\boldsymbol u}^{\prime}_{x^{\prime}}={\boldsymbol u}^{\prime}_{x}\frac{\partial x}{\partial x^{\prime}}+
{\boldsymbol u}^{\prime}_{t}\frac{\partial t}{\partial x^{\prime}}
\]
which gives
\[
{\boldsymbol u}^{\prime}_{x^{\prime}}=(b+{\boldsymbol r}^T{\boldsymbol u}^{\prime}){\boldsymbol u}^{\prime}_{x}
\]
and using the above form of ${\boldsymbol u}^{\prime}$ and the relations (\ref{sw:c10}) reduces to
\[
{\boldsymbol u}^{\prime}_{x^{\prime}}=\frac{{\boldsymbol u}^{\prime}_{x}}{a+{\boldsymbol p\cdot u}} .
\]
These expressions simplify (\ref{sw:c11}) to
\[
\Lambda {\boldsymbol u}^{\prime}_{x}=-\gamma ({\boldsymbol q}+A{\boldsymbol u})({\boldsymbol p}^TF{\boldsymbol u}_x)+\gamma (a+{\boldsymbol p\cdot u})AF{\boldsymbol u}_x .
\]
Using the relations (\ref{sw:c9}) and (\ref{sw:c10}) and differentiate the expression of ${\boldsymbol u}^{\prime}$
with respect to $x$ the above equation takes the form
\[
\Lambda \left [ (a+{\boldsymbol p}^T{\boldsymbol u})A{\boldsymbol u}_x-({\boldsymbol p\cdot u}_x)A{\boldsymbol u}-{\boldsymbol p u}_x{\boldsymbol q}\right ]=
-\gamma ({\boldsymbol q}+A{\boldsymbol u})({\boldsymbol p}^TF{\boldsymbol u}_x)+\gamma (a+{\boldsymbol p\cdot u})AF{\boldsymbol u}_x
\]
and simplifying to get
\[
\Lambda \left [ (a+{\boldsymbol p}^T{\boldsymbol u})A-(A{\boldsymbol u}+{\boldsymbol q}){\boldsymbol p}^T\right ]=
\gamma(a+{\boldsymbol p}^T{\boldsymbol u})^2\left [(a+{\boldsymbol p}^T{\boldsymbol u})A-(A{\boldsymbol u}+{\boldsymbol q}){\boldsymbol p}^T\right ]F .
\]
Solving for $F({\boldsymbol u})$ to obtain
\begin{gather}\label{sw:c12}
F({\boldsymbol u})=\frac{[H({\boldsymbol u})]^{-1}\Lambda H({\boldsymbol u})}{\gamma(a+{\boldsymbol p}^T{\boldsymbol u})^2},
\end{gather}
where
\[
H({\boldsymbol u})=(a+{\boldsymbol p}^T{\boldsymbol u})A-(A{\boldsymbol u}+{\boldsymbol q}){\boldsymbol p}^T .
\]

Summarizing we have:
\begin{theorem}\label{theorem1}
The nonlinear system of diffusion equations \eqref{sw:c5} can be mapped into the linear system
\eqref{sw:c6} by the equivalence transformation admitted by \eqref{sw:c5}
if and only if the functions $F({\boldsymbol u})$ is of the form \eqref{sw:c12}.
\end{theorem}

\begin{remark}
Point transformations that linearize system (\ref{sw:a3}) lead to contact transformations that linearize
system (\ref{sw:a1}).
\end{remark}

\begin{remark}
In the case where $\Lambda =I_2$, that is, system (\ref{sw:c1}) becomes two separate linear diffusion equations
with diffusivity constants equal to 1, the linearizing form of (\ref{sw:a1}) is
\begin{gather}\label{sw:c13}
u_t=\left [\frac {u_x}{(p_1u+p_2v+a)^2}\right ]_x,\qquad
v_t=\left [\frac {v_x}{(p_1u+p_2v+a)^2}\right ]_x .
\end{gather}
\end{remark}

It is known that the part of transformation (\ref{sw:cc1}), namely
$x^{\prime}=v$, $t^{\prime}=t$, $v^{\prime}=x$, which is known as {\it pure hodograph}
transformation maps the potential equation $v_t=v_x^{-2}v_{xx}$ into the linear heat equation
$v^{\prime}_{t^{\prime}}=v^{\prime}_{x^{\prime}x^{\prime}}$. In the spirit of the work in
\cite{sw:akhatov}, where such transformations were classified for the potential equation
$v_t=f(v_x)v_{xx}$, we consider the potential system of~(\ref{sw:a1}),
\begin{gather}
w_t=f(w_x,z_x)w_{xx}+g(w_x,z_x)z_{xx}, \nonumber \\
z_t=h(w_x,z_x)w_{xx}+k(w_x,z_x)z_{xx}. \label{sw:c14} 
\end{gather}
Transformations that presented in this section which linearize systems of the form (\ref{sw:a3}), can also
be employed to linearize systems of the form (\ref{sw:c14}). We present the results in the following theorem:

\begin{theorem}\label{theorem2}
The nonlinear system of potential diffusion equations \eqref{sw:c14} can be mapped into the linear system
\begin{gather*}
w^{\prime}_{t^{\prime}}=\mu_1w^{\prime}_{x^{\prime}x^{\prime}}+\mu_2z^{\prime}_{x^{\prime}x^{\prime}},  \qquad
z^{\prime}_{t^{\prime}}=\mu_3w^{\prime}_{x^{\prime}x^{\prime}}+\mu_4z^{\prime}_{x^{\prime}x^{\prime}}
\end{gather*}
by the transformation \eqref{sw:c3}
if and only if the functions $f(w_x,z_x)$, $g(w_x,z_x)$, $h(w_x,z_x)$, $k(w_x,z_x)$ are of the form
\[
\left [\begin{array}{cc}
f(w_x,z_x)&g(w_x,z_x)\\h(w_x,z_x)&k(w_x,z_x)
\end{array}\right ]=\frac{[H({\boldsymbol w}_x)]^{-1}\Lambda H({\boldsymbol w}_x)}
{\gamma(a+{\boldsymbol p}^T{\boldsymbol w}_x)^2},
\]
where
\[
H({\boldsymbol w}_x)=(a+{\boldsymbol p}^T{\boldsymbol w}_x)A-(A{\boldsymbol w}_x+{\boldsymbol q}){\boldsymbol p}^T .
\]
\end{theorem}

Using the above theorem the corresponding form of Remark~2 reads:
\begin{remark}
The nonlinear potential system
\[
w_t=\frac {w_{xx}}{(p_1w_x+p_2z_x+a)^2},\qquad
z_t=\frac {z_{xx}}{(p_1w_x+p_2z_x+a)^2}
\]
can be transformed into two separable linear heat equations.
\end{remark}

\section{Examples of linearizing mappings}

In this section we use the results of the previous section to present two examples.

\begin{example}
We consider the system (\ref{sw:c13}) with its corresponding second generating potential system
\begin{gather}
w_x=u,\qquad w_t=\frac {u_x}{(p_1u+p_2v+a)^2}, \nonumber \\
z_x=v,\qquad z_t=\frac {v_x}{(p_1u+p_2v+a)^2}. \label{sw:d1}
\end{gather}
We have shown that system (\ref{sw:d1}) can be linearized using the equivalence transformations of (\ref{sw:c1}).
That is, the transformation
\begin{gather*}
x^{\prime}=ax +p_1w+p_2z,\qquad t^{\prime}=t,\qquad w^{\prime}=q_1x+a_1w+a_2z,\qquad z^{\prime}=q_2x+a_3w+a_4z, \\
u^{\prime}=\frac{q_1+a_1u+a_2v}{a+p_1u+p_2v},\qquad
v^{\prime}=\frac{q_2+a_3u+a_4v}{a+p_1u+p_2v}
\end{gather*}
maps the linear system
\begin{gather*}
w^{\prime}_{x^{\prime}}=u^{\prime},\qquad w^{\prime}_{t^{\prime}}=u^{\prime}_{x^{\prime}}, \qquad
z^{\prime}_{x^{\prime}}=v^{\prime},\qquad z^{\prime}_{t^{\prime}}=v^{\prime}_{x^{\prime}}
\end{gather*}
into the nonlinear system (\ref{sw:d1}).

This transformation leads to the contact transformation
\begin{gather*}
{\rm d}x^{\prime}=a{\rm d}x+(u+v){\rm d}x+\left [(p_1u+p_2v+a)^{-2}(u_x+v_x)\right ]{\rm d}t,
\qquad {\rm d}t^{\prime}={\rm d}t, \\
u^{\prime}=\frac{q_1+a_1u+a_2v}{a+p_1u+p_2v},\qquad 
v^{\prime}=\frac{q_2+a_3u+a_4v}{a+p_1u+p_2v}
\end{gather*}
that maps the two separate linear diffusion equations
\[
u^{\prime}_{t^{\prime}}=u^{\prime}_{x^{\prime}x^{\prime}},\qquad
v^{\prime}_{t^{\prime}}=v^{\prime}_{x^{\prime}x^{\prime}}
\]
into the nonlinear system (\ref{sw:c13}). Using Theorem~\ref{theorem2} we deduce that the potential
form of (\ref{sw:c13}),
\[
w_t=\frac {w_{xx}}{(p_1w_x+p_2z_x+a)^2},\qquad
z_t=\frac {z_{xx}}{(p_1w_x+p_2z_x+a)^2}
\]
can be linearized by the transformation
\[
x^{\prime}=ax +p_1w+p_2z,\qquad
t^{\prime}=t,\qquad w^{\prime}=q_1x+a_1w+a_2z,\qquad z^{\prime}=q_2x+a_3w+a_4z.
\]
\end{example}

\begin{example}
We consider the special case of the equivalence transformation of (\ref{sw:a3})
\[
x^{\prime}=w+z,\qquad t^{\prime}=2t,\qquad u^{\prime}=\frac{v+1}{u+v},\qquad v^{\prime}=\frac{u+1}{u+v},\qquad 
w^{\prime}=x+z,\qquad z^{\prime}=x+w
\]
which maps the linear system
\begin{gather*}
w^{\prime}_{x^{\prime}}=u^{\prime},\qquad w^{\prime}_{t^{\prime}}=\mu_1u^{\prime}_{x^{\prime}}+\mu_2v^{\prime}_
{x^{\prime}},\qquad
z^{\prime}_{x^{\prime}}=v^{\prime},\qquad z^{\prime}_{t^{\prime}}=\mu_3u^{\prime}_{x^{\prime}}+\mu_4v^{\prime}_
{x^{\prime}}
\end{gather*}
into the nonlinear system
\begin{gather*}
w_x=u,\qquad w_t=\frac{\nu_1uv+\nu_2u+\nu_3v+\nu_4}{(u+v)^3}u_x+\frac{-\nu_1u^2+(\nu_2-\nu_3)u+\nu_4}{(u+v)^3}v_x, \\
z_x=v,\qquad z_t=\frac{\nu_1v^2+(\nu_2-\nu_3)v-\nu_4}{(u+v)^3}u_x+\frac{-\nu_1uv+\nu_3u+\nu_2v-\nu_4}{(u+v)^3}v_x,
\end{gather*}
where
\begin{gather*}
\nu_1=\mu_1-\mu_2+\mu_3-\mu_4,\qquad
\nu_2=\mu_1+\mu_2+\mu_3+\mu_4,\qquad
\nu_3=\mu_1-\mu_2-\mu_3+\mu_4,\\
\nu_1=\mu_1+\mu_2-\mu_3-\mu_4.
\end{gather*}
Clearly if we set $\mu_1=\mu_4=1$, $\mu_2=\mu_3=0$, we obtain a special case of the previous example.

Now the above transformation lead to the contact transformation
\begin{gather*}
{\rm d}x^{\prime}=(u+v){\rm d}x+\left [\nu_1(vu_x-uv_x)+\nu_2(u_x+v_x)\right ](u+v)^{-2}{\rm d}t,
\qquad {\rm d}t^{\prime}=2{\rm d}t, \\
u^{\prime}=\frac{1+v}{u+v},\qquad 
v^{\prime}=\frac{1+u}{u+v}
\end{gather*}
that maps the linear system
\[
u^{\prime}_{t^{\prime}}=\mu_1u^{\prime}_{x^{\prime}x^{\prime}}+\mu_2v^{\prime}_{x^{\prime}x^{\prime}},\qquad
v^{\prime}_{t^{\prime}}=\mu_3u^{\prime}_{x^{\prime}x^{\prime}}+\mu_4v^{\prime}_{x^{\prime}x^{\prime}}
\]
into the nonlinear system
\begin{gather*}
u_t=\left [\frac{\nu_1uv+\nu_2u+\nu_3v+\nu_4}{(u+v)^3}u_x+\frac{-\nu_1u^2+(\nu_2-\nu_3)u+\nu_4}
{(u+v)^3}v_x\right ]_x, \\
v_t=\left [\frac{\nu_1v^2+(\nu_2-\nu_3)v-\nu_4}{(u+v)^3}u_x+\frac{-\nu_1uv+\nu_3u+\nu_2v-\nu_4}{(u+v)^3}v_x\right ]_x .
\end{gather*}
\end{example}

\section{A linearizing case of the system (\ref{sw:a2})}

In Section 3 we used the equivalence transformations of (\ref{sw:a3}) to derive linearizing mappings.
We point out that employment of the equivalence transformations of (\ref{sw:a2}) does not lead to linea\-rizing
mappings. However such linea\-rizing mapping, which is not member of the equivalence transformations, exists for
a special case of (\ref{sw:a2}).

We consider the special case of (\ref{sw:a2})
\begin{gather}
w_x=u ,\qquad
w_t=-u^{-2}u_x,  \qquad
v_t=\left [u^{-2}v_x\right ]_x \label{sw:e1}
\end{gather}
which is the first generation potential system of
\begin{gather}
u_t=-\left [u^{-2}u_x\right ]_x , \qquad
v_t=\left [u^{-2}v_x\right ]_x .\label{sw:e2}
\end{gather}
System (\ref{sw:e1}) admits the infinite-dimensional Lie symmetries
\[
\Gamma_{\phi}=\phi(t,w)\frac{\partial~}{\partial x}-u^2\phi_w\frac{\partial~}{\partial u},\qquad 
\Gamma_{\psi}=\psi(t,w)\frac{\partial~}{\partial v},
\]
where the function $\phi(t,w)$ satisfies the backward linear heat equation $\phi_t+\phi_{ww}=0$
and $\psi(t,w)$ satisfies the linear heat equation $\psi_t-\psi_{ww}=0$. These symmetries induce
potential symmetries for the system (\ref{sw:e2}).

If a nonlinear PDE (or a system of PDEs) admits infinite-parameter groups, then it can be transformed
into a linear PDE (or into a linear system of PDEs) if these groups satisfy certain criteria. These criteria
and the method for finding the linearizing mapping using the infinite-dimensional symmetries can be found
in~\cite{sw:bluman}. Hence, using the method described in~\cite{sw:bluman}, the above infinite-dimensional Lie
symmetries of (\ref{sw:e1}) lead to the transformation
\[
x^{\prime}=w,\qquad t^{\prime}=t,\qquad u^{\prime}=\frac 1u,\qquad v^{\prime}=v,\qquad w^{\prime}=x
\]
which maps the linear system
\begin{gather*}
w^{\prime}_{x^{\prime}}=u^{\prime} , \qquad
w^{\prime}_{t^{\prime}}=-u^{\prime}_{x^{\prime}}, \qquad
v^{\prime}_{t^{\prime}}=v^{\prime}_{x^{\prime}x^{\prime}}
\end{gather*}
into the nonlinear system (\ref{sw:e1}). Consequently this mapping lead to the contact transformation
\[
{\rm d}x^{\prime}=u{\rm d}x+u^{-2}u_x{\rm d} t,\qquad
{\rm d}t^{\prime}={\rm d}t,\qquad
u^{\prime}=\frac 1u,\qquad
v^{\prime}=v
\]
which maps the linear separable system
\begin{gather*}
u^{\prime}_{t^{\prime}}=-u^{\prime}_{x^{\prime}x^{\prime}}, \qquad
v^{\prime}_{t^{\prime}}=v^{\prime}_{x^{\prime}x^{\prime}}
\end{gather*}
into the nonlinear system (\ref{sw:e2}).

We point out that the above result cannot be achieved using the equivalence transformations of (\ref{sw:a3}).
However it is straight forward to derive the above contact transformation as a special case of the one
derived in the Example~1, Section~4, which maps the two linear separable diffusion equations
$u^{\prime}_{t^{\prime}}=u^{\prime}_{x^{\prime}x^{\prime}}$ and
$v^{\prime}_{t^{\prime}}=v^{\prime}_{x^{\prime}x^{\prime}}$ into the nonlinear system
$u_t=\big [u^{-2}u_x\big ]_x$ and $v_t=\big [u^{-2}v_x\big]_x$.

\section{Conclusion}

In this paper we have considered the problem of finding forms of the general class of systems of
diffusion equations (\ref{sw:a1}) that can be linearized. We have employed the second generation potential
system (\ref{sw:a3}) and derived the equivalence transformations admitted by this system. Using these
transformations we classified special cases of (\ref{sw:a3}) that can be linearized. In turn these transformations
lead to nonlocal mappings that linearize the corresponding forms of (\ref{sw:a1}). Furthermore we determined
a special case of (\ref{sw:a1}) that can be linearized, by considering the Lie symmetries of the first generation
potential system (\ref{sw:a2}), which induce potential symmetries for the system (\ref{sw:a1}).

The question that arises here is: Are these the only cases of the system (\ref{sw:a1}) that can
be linea\-rized? Furthermore: Can the results obtained here be generalized for systems of $n$ equations?
These are two of the problems of our investigation in the near future.

\subsection*{Acknowledgements}
Both authors wish to acknowledge the financial support of this project by their two Universities.

\LastPageEnding


\begin{thebibliography}{99}

\footnotesize

\bibitem{sw:akhatov}
Akhatov I.Sh., Gazizov R.K., Ibragimov N.Kh., Nonlocal symmetries. Heuristic approach,
{\it J. Soviet. Math.}, 1991, V.55, 1401--1450.

\bibitem{sw:ron1}
Baikov V.A, Gladkov A.V., Wiltshire R.J., Lie symmetry
classification analysis for nonlinear coupled diffusion,  {\it J.
Phys. A: Math. Gen.}, 1998, V.31, 7483--7499.

\bibitem{sw:bluman}
Bluman G.W., Kumei S., Symmetries and differential equations, New York, Springer, 1989.

\bibitem{sw:bluman2}
Bluman G.W., Kumei S., On the remarkable nonlinear diffusion equation
$(\partial /\partial x)[a(u+b)^{-2}(\partial u /\partial x)]-(\partial u/\partial t)=0$,
{\it J. Math. Phys.}, 1980, V.21,  1019--1023.

\bibitem{sw:ibragimov}
Ibragimov N.H., Torrisi M., Valenti A., Preliminary group classification of equations
$v_{tt}=f(x,v_x)v_{xx}+g(x,v_x)$, {\it J. Math. Phys.}, 1991, V.32,  2988--2995.


\bibitem{sw:jury}
Jury W.A., Letey J., Stolzy L.H., Flow of water and energy
under desert conditions,  in Water in Desert Ecosystems, Editors D.~Evans and J.L.~Thames,
 Stroudsburg, PA: Dowden, Hutchinson and
Ross, 1981, 92--113.

\bibitem{sw:ovsiannikov}
Ovsiannikov L.V.,  Group analysis of differential equations, New York, Academic, 1982.

\bibitem{sw:philip}
Philip J.R., de Vries D.A., Moisture movement in porous media
under temperature gradients, {\it Trans. Am. Geophys. Un.}, 1957,
V.38, 222--232.

\bibitem{sw:sw}
Sophocleous C., Wiltshire R.J., Systems of diffusion equations, in Proceedings of 11th
Conference ``Symmetry
in Physics'', Prague,  2004, 17~pages,

\bibitem{sw:ron2}
Wiltshire R.J., The use of Lie transformation groups in the
solution of the coupled diffusion equation, {\it J.~Phys. A: Math.
Gen.}, 1994, V.27, 7821--7829.

\end{thebibliography}
\end{document}